\documentclass[conference]{IEEEtran}
\IEEEoverridecommandlockouts
\usepackage[htt]{hyphenat}
\usepackage[hyphenbreaks]{breakurl}
\usepackage[hyphens]{url}
\usepackage{booktabs}
\usepackage{tabularx}
\usepackage{tikz}
\usepackage{pgfplots}
\usepackage{pgfplotstable}
\usepgfplotslibrary{
        groupplots,
        statistics,
    }
\pgfplotsset{select coords between index/.style 2 args={
    x filter/.code={
        \ifnum\coordindex<#1\fi
        \ifnum\coordindex>#2\fi
    }
}}

\usepackage{pgfplotstable}
\pgfplotstableset{
col sep=semicolon
}

\usepackage{amsmath,amssymb,amsfonts}
\usepackage{algorithmic}
\usepackage{graphicx}
\usepackage{textcomp}
\usepackage{xcolor}
\def\BibTeX{{\rm B\kern-.05em{\sc i\kern-.025em b}\kern-.08em
    T\kern-.1667em\lower.7ex\hbox{E}\kern-.125emX}}

\IEEEoverridecommandlockouts

\usepackage{tikz}
\newcommand\copyrighttext{%
  \footnotesize 
  Copyright \textcopyright 2020 IEEE. Personal use of this material is permitted.
  Permission from IEEE must be obtained for all other uses, in any current or future 
  media, including reprinting/republishing this material for advertising or promotional 
  purposes, creating new collective works, for resale or redistribution to servers or 
  lists, or reuse of any copyrighted component of this work in other works. 
  DOI: will follow as soon as the paper is published.
  }
\newcommand\copyrightnotice{%
\begin{tikzpicture}[remember picture,overlay]
\node[anchor=south,yshift=10pt] at (current page.south) {\fbox{\parbox{\dimexpr\textwidth-\fboxsep-\fboxrule\relax}{\copyrighttext}}};
\end{tikzpicture}%
}   
    
\begin{document}

\title{How Fast Can We Insert? An Empirical Performance Evaluation of Apache Kafka
}

\author{\IEEEauthorblockN{Guenter Hesse, Christoph Matthies, Matthias Uflacker}
\IEEEauthorblockA{\textit{Hasso Plattner Institute} \\
\textit{University of Potsdam}\\
Germany\\
firstname.lastname@hpi.de}
}

\maketitle

\copyrightnotice

\begin{abstract}
Message brokers see widespread adoption in modern IT landscapes, with Apache Kafka being one of the most employed platforms.
  These systems feature well-defined APIs for use and configuration and present flexible solutions for various data storage scenarios.
  Their ability to scale horizontally enables users to adapt to growing data volumes and changing environments.
  However, one of the main challenges concerning message brokers is the danger of them becoming a bottleneck within an IT architecture.
  To prevent this, knowledge about the amount of data a message broker using a specific configuration can handle needs to be available.
  In this paper, we propose a monitoring architecture for message brokers and similar Java Virtual Machine-based systems.
  We present a comprehensive performance analysis of the popular Apache Kafka platform using our approach.
  As part of the benchmark, we study selected data ingestion scenarios with respect to their maximum data ingestion rates.
  The results show that we can achieve an ingestion rate of about 420,000\,messages/second on the used commodity hardware and with the developed data sender tool.
\end{abstract}


\section{Introduction}
In the current business landscape, with an ever-increasing growth in data and popularity of cloud-based applications, horizontal scalability is becoming an increasingly common and important requirement.
Message brokers play a central role in modern IT systems as they satisfy this requirement and thus, allow for adaptations of the IT landscape to data sources that grow both in volume and velocity.
Moreover, they can be used to decouple disparate data sources from applications using this data.
Usage scenarios where message brokers are employed are manifold and reach from e.g., machine learning~\cite{DBLP:conf/kdd/ZhuangL19} to stream processing architectures~\cite{DBLP:conf/icpads/HesseL15,DBLP:conf/debs/HesseMRU17,DBLP:conf/tpctc/HesseRMLKU17} and general-purpose data processing~\cite{DBLP:journals/pvldb/ZhangWLXL17}.

In the context of a complex IT architecture, the degree, to which a system aligns with its application scenarios and the functional and non-functional requirements derived from it, are key~\cite{DBLP:conf/hicss/LorenzRHUP17}.
If non-functional requirements related to performance are not satisfied, the system might become a bottleneck.
This situation does not directly imply that the system itself is inadequate for the observed use case, but might indicate a suboptimal configuration.
Therefore, it is crucial to be able to evaluate the capabilities of a system in certain environments and with distinct configurations.
The knowledge about such study results is a prerequisite for making informed decisions about whether a system is suitable for the existing use cases.
Additionally, it is also crucial for finding or fine-tuning appropriate system configurations.

The contributions of this research are as follows:
\begin{itemize}
\item
We propose a user-centered and extensible monitoring framework, which includes tooling for analyzing any JVM-based system.
\item
We present an analysis that highlights the capabilities of Apache Kafka regarding the maximum achievable rate of incoming records per time unit.
\item
We enable reproducibility of the presented results by making all needed artifacts available online\footnote{\url{https://github.com/guenter-hesse/KafkaAnalysisTools}}.
\end{itemize}

The rest of the paper is structured as follows: In Section~\ref{sec:kafka} we give a brief introduction of Apache Kafka.
Section~\ref{sec:monarch} presents the benchmark setup and the developed data sender tool.
Subsequently, we describe the results of the ingestion rate analyses.
Section~\ref{chap:rw} introduces related work and Section~\ref{sec:lessonslearned} elaborates on the lessons learned.
The last chapter concludes the study and outlines areas of future work.

\section{Apache Kafka}
\label{sec:kafka}

Apache Kafka is a distributed open-source message broker or messaging system originally developed at \emph{LinkedIn} in 2010~\cite{kafkaLinkedin}.
The core of this publish-subscribe system is a distributed commit log, although it has extended its scope through extensions.
An example is Kafka Streams~\cite{kafkaStreams}, a client library for developing stream processing applications.

\begin{figure}[!htb]
\centering
\includegraphics[width=\columnwidth]{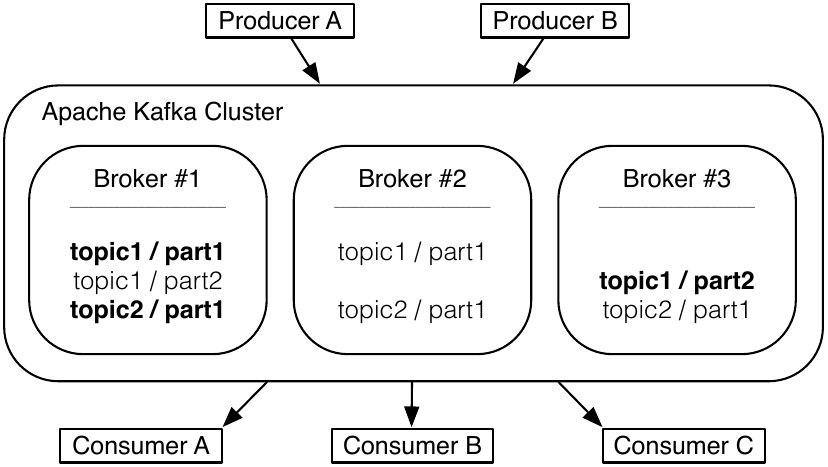}
\caption{Apache Kafka cluster architecture (based on~\cite{kreps2011kafka})}
\label{fig:kafkaarch}
\end{figure}

The high-level architecture of an exemplary Apache Kafka cluster is visualized in Figure~\ref{fig:kafkaarch}.
A cluster consists of multiple brokers, which are numbered and store data assigned to topics.
Data producers send data to a certain topic stored in the cluster.
Consumers subscribe to a topic and are forwarded new values sent to this topic as soon as they arrive.

Topics are divided into partitions.
The number of topic partitions can be configured at the time of topic creation.
Partitions of a single topic can be distributed across different brokers of a cluster.
However, a message order across partitions is not guaranteed by Apache Kafka~\cite{kafka,kreps2011kafka}.

Next to the number of partitions, it is possible to define a replication factor for each topic, one being the minimum.
This allows preventing data loss in the case of a single broker failure.
In the context of replication, Apache Kafka defines \emph{leaders} and \emph{followers} for each partition.
The leader handles all reads and writes for the corresponding topic partition, whereas followers copy or replicate the inserted data.
In Figure~\ref{fig:kafkaarch}, the leader partitions are shown in bold type.
The first topic, \textit{topic1} has two partitions and a replication factor of one, while \textit{topic2} has only one partition and a replication factor of two~\cite{kafka}.

\begin{figure}[!htb]
\centering
\includegraphics[width=0.9\columnwidth]{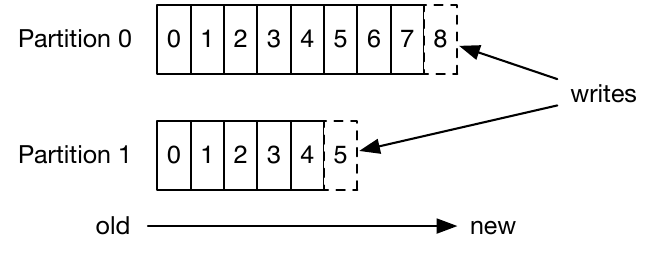}
\caption{Apache Kafka topic structure (based on~\cite{kafkaProducerConfig})}
\label{fig:kafkapartitions}
\end{figure}

Figure~\ref{fig:kafkapartitions} shows the structure of an Apache Kafka topic, specifically of a topic with two partitions.
Each of these partitions is an ordered and immutable record sequence where new values are appended.
A sequential number is assigned to each topic record within a partition, referred to as an \textit{offset}.
Apache Kafka itself provides the topic \textit{\_\_consumer\_offsets} for storing the offsets.
However, consumers must manage their \textit{offset}.
They can commit their current \textit{offset} either automatically in certain intervals or manually.
The latter can be done either synchronously or asynchronously.
When polling data, a consumer needs to pass the \textit{offset} to the cluster.
Apache Kafka returns all messages with a greater \textit{offset}, i.e., all new messages that have not already been sent to this consumer.
As the consumer has control over its \textit{offset}, it can also decide to start from the beginning and to reread messages~\cite{kafka}.

Furthermore, Apache Kafka can be configured to use the \textit{LogAppendTime} feature, which induces Apache Kafka to assign a timestamp to each message once it is appended to the log.
The existing alternative, which represents the default value, is \textit{CreateTime}.
In this setting, the timestamp created by the Apache Kafka producer when creating the message, i.e., before sending it, is stored along with the message.
For transmitting messages, a producer can require multiple retries, which would increase the difference between the timestamp assigned with a message and the time when it is appended to the log and thus, made available for consuming applications~\cite{kafka}.

\section{Benchmark Setup}
This section introduces the monitoring architecture employed in the ingestion rate study as well as the developed data sender tool.

\subsection{Monitoring Architecture}
\label{sec:monarch}

\begin{figure}[!htb]
\centering
\includegraphics[width=0.85\columnwidth]{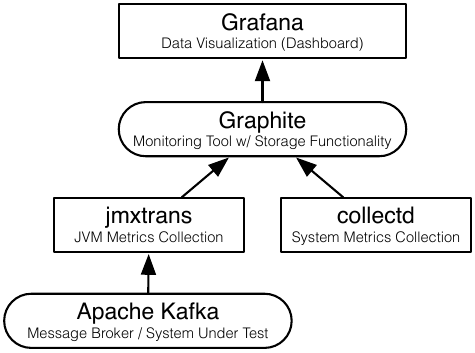}
\caption{Monitoring architecture in Fundamental Modeling Concepts (FMC)~\cite{knopfel2005fundamental}}
\label{fig:arch}
\end{figure}

\begin{sloppypar}
The architecture of the monitoring system is shown in Figure~\ref{fig:arch}.
We use \textit{Grafana}~\cite{grafana}, an open-source tool for creating and managing dashboards and exporting data, as the interface to the user.
The presented benchmarks employ version 5.4.5 of its \textit{docker} image.
OS-level virtualization through docker is used for ease of installation and replicability of results.
The OS base image used in this image allows a simple time zone configuration via an environment variable, which is important for time synchronization among all systems.
Later versions of the image contain a different OS, specifically \textit{Alpine Linux}~\cite{alpinelinux}, which no longer supports this feature.
\end{sloppypar}

\textit{Grafana} fetches the data to display from \textit{Graphite}~\cite{graphite}, an open-source monitoring tool.
It consists of three components: \textit{Carbon}, \textit{Whisper}, and \textit{Graphite-web}.
\textit{Carbon} is a service that retrieves time-series data, which is stored in \textit{Whisper}, a persistence library.
\textit{Graphite-web} includes an interface for designing dashboards.
However, these dashboards are not as appealing and functionally comprehensive as the corresponding components of \textit{Grafana}, which is why \textit{Grafana} is employed.
For the installation of \textit{Graphite}, the official \textit{docker} image in version 1.1.4 is used, again for time zone configuration reasons.

\textit{Graphite} receives its input from two sources: \textit{collectd}~\cite{collectd} and \textit{jmxtrans}~\cite{jmxtrans}.
The former is a daemon collecting system and application performance metrics, that runs on the broker's machines in the described setup.
It offers plugins for gathering OS-level measurements, such as memory usage, system load, and received or transmitted packages over the network.

\textit{Jmxtrans}, the other data source for \textit{Graphite}, is a tool for collecting JVM runtime metrics.
These metrics are provided via Java Management Extensions (JMX)~\cite{jmx}.
Using \textit{jmxtrans} we tracked internal metrics, such as JVM memory usage, the number of incoming bytes, and the number of messages entering Apache Kafka per time unit.

Apache Kafka is the system under test (SUT) in the evaluation of this paper.
It can be exchanged for any other system running in a JVM, i.e., the proposed architecture is not limited to Apache Kafka or message brokers in general.
The information gathered in \textit{Graphite} is summarized in a \textit{Grafana} dashboard.
Exports of the collected \textit{Grafana} data enable further, more detailed analysis.

\begin{table}[!htb]
\caption{Characteristics of the Apache Kafka broker nodes}
\begin{center}
\begin{tabularx}{\columnwidth}{@{}lX@{}}
\toprule
Characteristic & Value \\ \midrule
Operating system               &  Ubuntu 18.04.2 LTS    \\
    CPU           &   Intel(R) Xeon(R) CPU E5-2697 v3 @ 2.60GHz, 8 cores    \\
    RAM           &   32GB    \\
    Network bandwidth      & 1Gbit: \newline - measured bandwidth between nodes: 117.5\,MB/s \newline - measured bandwidth of intra-node transfer: 908\,MB/s\\
    Disk & min. 13 Seagate ST320004CLAR2000 in RAID 6, access via Fibre Channel with 8Gbit/s: \newline measured write performance about 70\,MB/s \\
    Hypervisor & VMware ESXi 6.7.0\\
    Kafka version           &   2.3.0    \\
    Java version & OpenJDK 1.8.0\_222 \\ \bottomrule
\end{tabularx}
\end{center}
\label{tab:system}
\end{table}

Apache Kafka is installed on three virtual machines featuring identical hardware setups and configurations, which are shown in Table~\ref{tab:system}.
We use a commodity network setup whose bandwidth we determined using \textit{ipfer3}~\cite{iperf}.
The write performance is measured using the Unix command-line tool \textit{dd}~\cite{gnudd}, specifically with the following command: ``\texttt{dd if=/dev/zero of=/opt/kafka/test1.img bs=1G count=1 oflag=dsync}''.
The data sender is a Scala application compiled to a fat \textit{jar} file and executed using OpenJDK 1.8 with the default parallel garbage collector (\emph{ParallelGC}).
The data sender is assigned an initial memory allocation pool of 1\,GB while the maximum size of this pool is about 14\,GB.
Apache Kafka uses an initial and maximum memory allocation pool of 1\,GB and the \textit{Garbage-First garbage collector} (G1 GC).
Additional arguments passed to the Apache Kafka JVMs are \texttt{MaxGCPauseMillis=20}, \texttt{InitiatingHeapOccupanyPercent=35}, as well as \texttt{ExplicitGCInvokesConcurrent}, which fine-tune the garbage collection behavior.

\subsection{Data Sender}
\label{sec:datasenderconfig}

To study the attainable ingestion rates of Apache Kafka, we developed a configurable data sender tool in the Scala programming language, which is part of the published artifacts.
It uses the Apache Kafka producer class for sending data to the message broker.

Table~\ref{tab:kafka} shows the default configuration parameters, i.e., \emph{properties}, that the data sender applies to the Apache Kafka producer.
Unless otherwise stated, these are the parameters employed in the presented measurements in Section~\ref{sec:ingestionrate}.
An Apache Kafka producer batches messages to lower the number of requests, thereby increasing throughput.
The \emph{batch-size} property limits the size of these message packages.
The used value is the default of 16,384\,bytes, as defined in the Apache Kafka documentation~\cite{kafkaProducerConfig}.
The \emph{acks} producer property determines the level of acknowledgments for sent messages.
There are three different options for the \emph{acks} configuration:
\begin{itemize}
  \item \textit{0}: The producer does not wait for any acknowledgment and counts the message as sent as soon as it is added to the socket buffer.
  \item \textit{1}: The leader will send an acknowledgment to the producer as soon as the message is written to its local log.
  The leader will not wait until its followers, i.e., other brokers, have written it to their log.
  \item \textit{all}: The leader waits until all in-sync replicas acknowledge the message before sending an acknowledgment to the producer.
  By default, the minimum number of in-sync replicas is set to one.
\end{itemize}

\begin{table}[]
\caption{Apache Kafka default producer properties}
\begin{center}
\begin{tabularx}{\columnwidth}{@{}lX@{}}
\toprule
Property & Value \\ \midrule
key-serializer-class & org.apache.kafka.common.serialization.StringSerial-izer \\
value-serializer-class & org.apache.kafka.common.serialization.StringSerial-izer \\
batch-size & 16,384\,bytes \\
buffer-memory-size & 33,554,432\,bytes \\
acks & 0 \\
\bottomrule
\end{tabularx}
\end{center}
\label{tab:kafka}
\end{table}

In addition to the configuration of the Apache Kafka producer, the developed data sender tool can be customized.
The \textit{read-in-ram} Boolean setting determines how inputs are read.
If \textit{read-in-ram} is not set, the data source object returns an iterator object of the records.
If it is set, the source object first loads the entire data set into memory by converting it into a list and then returns an iterator object for the created data structure.
Unless otherwise stated, \textit{read-in-ram} is enabled in the presented results.
The number of messages the data sender emits per time unit can be controlled using the \texttt{java.util.} \texttt{concurrent.ScheduledThreadPoolExecutor} class.
It can execute a thread periodically by applying a configurable delay.
Using this parameter, we can determine how many messages are to be sent per time unit.
Each execution sends a single message to Apache Kafka.
A configured delay of, e.g., 10K\,ns, leads to an input rate of 100K\,messages/second (MPS).

\section{Ingestion Rate Analysis}
\label{sec:ingestionrate}

This section presents the Apache Kafka ingestion rate analysis, starting with a description of the benchmark process.
It comprises analyzing three selected input rates with varying configurations regarding \textit{acks} levels, \textit{batch size}, data sender locality, \textit{read-in-ram} option, and data sender processes.

\subsection{Benchmark Execution Process}
Each analysis run lasts ten minutes.
The main characteristic studied is the number of incoming or ingested messages, particularly, the one-minute rate of this key performance indicator (KPI), i.e., the number of incoming messages during the last minute.
If not stated otherwise, the data sender is executed on the broker server where the topic is stored.

To reduce the number of manual steps needed, \textit{Ansible}~\cite{ansible} is used for automation.
Starting the \textit{Ansible} script triggers a build of the data sender project, the creation of a topic, and the assignment of this topic to the first of our three Apache Kafka brokers.
For all measurements, we use topics with a single partition and a replication factor of one.
Having one partition is a setting used for scenarios in which the order of data is crucial.
That is the case as Apache Kafka only makes guarantees for the correct message order within a partition, as outlined in Section~\ref{sec:kafka}.

After the Apache Kafka topics are prepared, the data sender is started.
Subsequently, a rise in the number of incoming messages of Apache Kafka can be observed using the \textit{Grafana} dashboard.
Once the configured send period is over, the \textit{Ansible} script stops and the dashboard charts adapt correspondingly.
The dashboard data is then exported as CSV.
The timeframe of these exports is configurable in \textit{Grafana}.

We incorporate the data set of the Grand Challenge published 2012 at the conference Distributed and Event-Based Systems (DEBS)~\cite{DBLP:conf/debs/JerzakHFGHS12} as input.
It contains data captured from multiple sensors that are combined into single records by an embedded PC within the manufacturing equipment.
One record comprises 66 columns with numerical and Boolean values.
When the end of the input file is reached, the data sender starts again from the beginning.

\subsection{Result Overview}
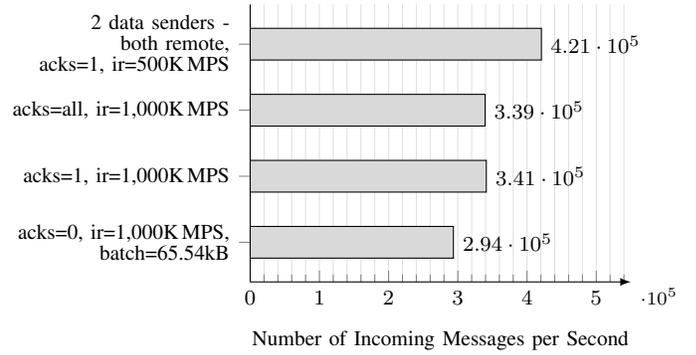
\begin{figure}[h]
\centering
\begin{tikzpicture}
\begin{axis}[
    xlabel=Number of Incoming Messages per Second,
    every x tick scale label/.style={at={(xticklabel cs:1)},anchor=south west,
    font=\fontsize{7}{7}\selectfont},
    y tick label style={
    rotate=0,
    font=\fontsize{7}{7}\selectfont,
    anchor=east,
    align=right},
    ytick=data,
    yticklabels={
    {acks=0, ir=1,000K\,MPS,\\batch=65.54kB},
    {acks=1, ir=1,000K\,MPS},
    {acks=all, ir=1,000K\,MPS},
    {2 data senders -\\both remote,\\acks=1, ir=500K\,MPS}},
    axis y line*=left,
    axis x line*=bottom,
    enlarge y limits = 0.20,
    x axis line style = {-latex},
    xbar,
    nodes near coords,
    nodes near coords align={horizontal},
    every node near coord/.append style={font=\fontsize{8}{8}\selectfont},
    bar width=12pt,
    bar shift=0pt,
    minor x tick num=4,
    xmajorgrids=true,
    xminorgrids=true,
    grid style={line width=.1pt, draw=gray!25},
    xmin=0,
    xmax=550000,
    width=\columnwidth*0.75,
    ymax=3,
    x tick label style={font=\fontsize{8}{8}\selectfont},
    x label style={font=\fontsize{8}{8}\selectfont},
    y tick label style={font=\fontsize{8}{8}\selectfont},
    y label style={font=\fontsize{8}{8}\selectfont},
    height=15em,
]

\addplot [
fill=gray!30
] table [
x=Value,
y expr=\coordindex,
col sep=tab
]
{perf_analysis/dfSummary.csv};

\end{axis}
\end{tikzpicture}
\caption{Ingested messages/second - one-minute rate}
\label{fig:summary}
\end{figure}

Figure~\ref{fig:summary} shows the maximum achieved input rates (\emph{ir}) of Apache Kafka for the selected configurations.
The input rates illustrated in all figures are the one-minute rates of incoming MPS, which is a KPI provided by Apache Kafka.
For all benchmark scenarios with the maximum configured input of 1,000K\,MPS, we selected the runs with the most stable input rates.

The highest input rate with about 421K\,MPS was achieved with two distinct data sender processes, each sending 250K\,MPS.
However, this is less than the configured input rate.
With a single data sender configured to send 1,000K\,MPS, the input rates are lower.
The results for the \textit{acks} levels of \textit{1} and \textit{all} are similar with input rates around 340K\,MPS.
Surprisingly, sending messages without waiting for acknowledgment, i.e., \textit{acks} set to \textit{0}, decreased the achieved input rate.
The maximum is at about 294K\,MPS with increased batch size.
In contrast to the other benchmark scenarios, the achievable input rate with acknowledgments disabled could be positively influenced by a higher batch size without harming the stability of the input rate.

\subsection{Input Rate of 100,000 Messages/Second}
\begin{figure}[!htb]
\centering
\begin{tikzpicture}
\begin{axis}[
    xlabel=Passed Time in Seconds,
    ylabel=Incoming Messages/Second,
    y label style={at={(axis description cs:0.04,.5)}},
    axis y line*=left,
    axis x line*=bottom,
    axis lines = left,
    minor tick num=4,
    grid=both,
    grid style={line width=.1pt, draw=gray!25},
    xmin=0,
    xmax=620,
    ymin=0,
    width=0.95\columnwidth,
     legend style=
    {at={(0.62,0.02)},
    anchor=south,
    draw=none,
    font=\fontsize{8}{8}\selectfont
    },
    x tick label style={font=\fontsize{8}{8}\selectfont},
    x label style={font=\fontsize{8}{8}\selectfont},
    y tick label style={font=\fontsize{8}{8}\selectfont},
    y label style={font=\fontsize{8}{8}\selectfont},
    height=17em
]

\addplot[mark=none, black,dashed, thick, samples=2] coordinates {(0,100000) (615,100000)};

\addplot[
color = blue,
mark=x,
mark size=1pt
] table [
y=Value,
x=Time,
col sep=tab
]
{perf_analysis/df3.csv};

\addplot[
color = green,
mark=x,
mark size=1pt
] table [
y=Value,
x=Time,
col sep=tab
]
{perf_analysis/df4.csv};

\addplot[
color = purple,
mark=x,
mark size=1pt
] table [
y=Value,
x=Time,
col sep=tab
]
{perf_analysis/df5.csv};

\addplot[
color = brown,
mark=x,
mark size=1pt
] table [
y=Value,
x=Time,
col sep=tab
]
{perf_analysis/df6.csv};

\addplot[
color = gray,
mark=x,
mark size=1pt
] table [
y=Value,
x=Time,
col sep=tab
]
{perf_analysis/df15.csv};

\legend{
100K\,MPS,
remote; acks=0; read-in-ram=false,
local; acks=0; read-in-ram=false,
local; acks=0; read-in-ram=true,
local; acks=1; read-in-ram=true,
local; acks=all; read-in-ram=true
}
\end{axis}
\end{tikzpicture}
\caption{Ingested messages/second - one-minute rate, configured 100K\,MPS}
\label{fig:100k}
\end{figure}
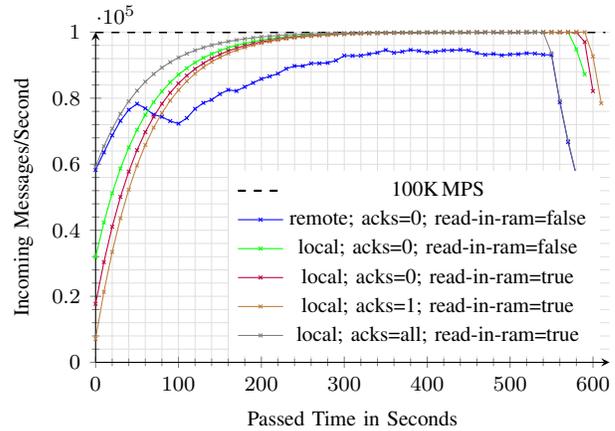

Figure~\ref{fig:100k} visualizes the one-minute rate of incoming MPS for a configured input of 100K\,MPS.
The parameters under investigation for this benchmark series are the data sender locality, the \textit{acks} level, and the \textit{read-in-ram} option.
Similar to all other observations, an increase in the number of incoming MPS can be seen at the beginning.
This is when the data sender is started and the one-minute rate begins to adapt accordingly.
Also, a sudden decrease in ingested messages is present in all charts after the data sender has transmitted messages for the configured duration and has shut down.
Consequently, the most interesting part of the evaluations is the data presented in the center of plots.

Figure~\ref{fig:100k} further shows that almost all chosen settings reach the configured input of 100K\,MPS.
The only exception is the \emph{remote; acks=0; read-in-ram=false} configuration, which is represented by the blue line.
In this setting, the data sender was executed remotely, specifically on a 2015 Apple MacBook Pro, which was connected using Ethernet.
For the other benchmark runs, the data sender was executed on the broker where the topic is stored.
All of the tested \textit{acks} level and \textit{read-in-ram} combinations reach the configured ingestion rate of 100K\,MPS.

\subsection{Input Rate of 250,000 Messages/Second}
\begin{figure}[!htb]
\centering
\begin{tikzpicture}
\begin{axis}[
    xlabel=Passed Time in Seconds,
    ylabel=Incoming Messages/Second,
    y label style={at={(axis description cs:0.04,.5)}},
    axis y line*=left,
    axis x line*=bottom,
    axis lines = left,
    minor tick num=4,
    grid=both,
    grid style={line width=.1pt, draw=gray!25},
    xmin=0,
    xmax=620,
    ymin=0,
    width=0.95\columnwidth,
    legend pos=south east,
    x tick label style={font=\fontsize{8}{8}\selectfont},
    x label style={font=\fontsize{8}{8}\selectfont},
    y tick label style={font=\fontsize{8}{8}\selectfont},
    y label style={font=\fontsize{8}{8}\selectfont},
    height=15em,
        legend style=
    {at={(0.6,0.02)},
    anchor=south,
    draw=none,
    font=\fontsize{8}{8}\selectfont
    },
       ]
 
 \addplot[mark=none, black,dashed, thick, samples=2] coordinates {(0,250000) (615, 250000)};
       
\addplot[
color = blue,
mark=x,
mark size=1pt
] table [
y=Value,
x=Time,
col sep=tab
]
{perf_analysis/df11.csv};

\addplot[
color = green,
mark=x,
mark size=1pt
] table [
y=Value,
x=Time,
col sep=tab
]
{perf_analysis/df12.csv};

\addplot[
color = purple,
mark=x,
mark size=1pt
] table [
y=Value,
x=Time,
col sep=tab
]
{perf_analysis/df13.csv};

\addplot[
color = brown,
mark=x,
mark size=1pt
] table [
y=Value,
x=Time,
col sep=tab
]
{perf_analysis/df14.csv};

\legend{
250K\,MPS,
acks=0; read-in-ram=false,
acks=0; read-in-ram=true,
acks=1; read-in-ram=true,
acks=all; read-in-ram=true
}
\end{axis}
\end{tikzpicture}
\caption{Ingested messages/second - one-minute rate, configured 250K\,MPS}
\label{fig:250k}
\end{figure}
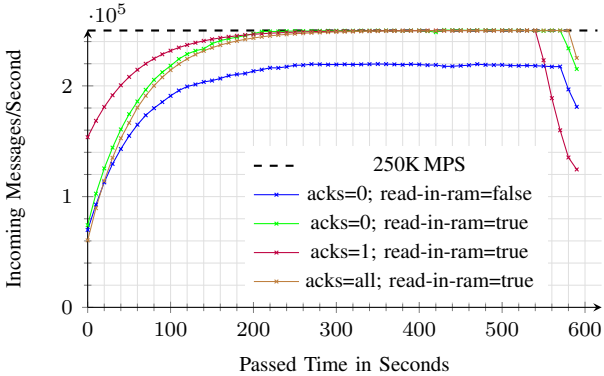

Figure~\ref{fig:250k} shows the results for an input rate of 250K\,MPS.
As we already identified the limits of the commodity hardware in Figure~\ref{fig:100k}, we do not pursue further tests with the laptop configuration.
Figure~\ref{fig:250k} highlights the significance of the \textit{read-in-ram} configuration, detailed in Section~\ref{sec:datasenderconfig}.
Particularly, the three configurations where \textit{read-in-ram} is set to \textit{true} reach the configured input of 250K\,MPS, whereas the run where \textit{read-in-ram} is set to \textit{false}, represented in blue, does not.
This configuration, where \textit{read-in-ram} is not active, is not able to handle more than about 220K\,MPS.
Thus, enabling \textit{read-in-ram} has a positive influence on the achievable number of incoming messages per second as the latency for accessing the main memory is lower than for accessing the disk.
It is evident that with \textit{read-in-ram} disabled, there is a bottleneck at the data sender side at this configured input rate.
The data can not be read as fast as it is required to achieve an ingestion rate of 250K\,MPS.

\subsection{Input Rate of 1,000K\,Messages/Second}
\begin{figure}[]
\centering
\begin{tikzpicture}
\begin{axis}[
    xlabel=Passed Time in Seconds,
    ylabel=Incoming Messages/Second,
    y label style={at={(axis description cs:0.04,.5)}},
    axis y line*=left,
    axis x line*=bottom,
    axis lines = left,
    xmin=0,
    xmax=620,
    ymin=0,
    grid=both,
    grid style={line width=.1pt, draw=gray!25},
    minor tick num=4,
    width=0.95\columnwidth,
    legend pos=south east,
    x tick label style={font=\fontsize{8}{8}\selectfont},
    x label style={font=\fontsize{8}{8}\selectfont},
    y tick label style={font=\fontsize{8}{8}\selectfont},
    y label style={font=\fontsize{8}{8}\selectfont},
    height=18em,
    legend columns=2,
     legend style=
    {at={(0.5,0.45)},
    anchor=south,
    draw=none,
    font=\fontsize{8}{8}\selectfont,
    cells={align=left}
    },
]

\addplot[mark=none, black,dashed, thick, samples=2] coordinates {(0,1000000) (615,1000000)};

\addplot[
color = green,
mark=x,
mark size=1pt
] table [
y=Value,
x=Time,
col sep=tab
]
{perf_analysis/df8.csv};

\addplot[
color = brown,
mark=x,
mark size=1pt
] table [
y=Value,
x=Time,
col sep=tab
]
{perf_analysis/df19.csv};

\addplot[
color = violet,
mark=x,
mark size=1pt
] table [
y=Value,
x=Time,
col sep=tab
]
{perf_analysis/df9.csv};

\addplot[
color = blue,
mark=x,
mark size=1pt
] table [
y=Value,
x=Time,
col sep=tab
]
{perf_analysis/df20.csv};

\addplot[
color = cyan,
mark=x,
mark size=1pt
] table [
y=Value,
x=Time,
col sep=tab
]
{perf_analysis/df10.csv};

\addplot[
color = red,
mark=x,
mark size=1pt
] table [
y=Value,
x=Time,
col sep=tab
]
{perf_analysis/df21.csv};

\legend{
1{,}000K\,MPS,
acks=0;\\batch=16.38\,kB,
acks=0; batch=65.54\,kB,
acks=1;\\batch=16.38\,kB,
acks=1; batch=65.54\,kB,
acks=all;\\batch=16.38\,kB,
acks=all; batch=65.54\,kB
}
\end{axis}
\end{tikzpicture}
\caption{Ingested messages/second - one-minute rate, configured 1,000K\,MPS}
\label{fig:1mioack0}
\end{figure}
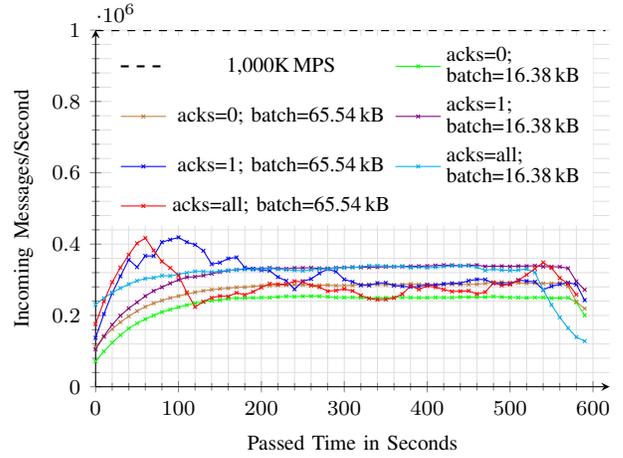

Figure~\ref{fig:1mioack0} visualizes the results for an input rate of 1,000K\,MPS.
As we discovered the limits of configurations with \textit{read-in-ram} set to \textit{false} previously, the parameter is enabled for all following measurements.
Next to testing different \textit{acks} levels, we analyze the effects of changes to the \textit{batch size}.
Particularly, we study the default size and a \textit{batch size} increased by a factor of four, which results in 65.54\,kB.

None of the configurations reach the configured ingestion rate.
While the highest ingestion rates peak at about 420K\,MPS for a short period, the lowest one is at about 250K\,MPS.
The two configurations that achieve this maximum peak are the ones with a \textit{batch size} of 65.5\,4kB and \textit{acks} set to \textit{1} and \textit{all}.
However, these are also the only two scenarios where no steady ingestion rate could be established.
The \textit{acks} level of \textit{0} combined with the default \textit{batch size} reached the lowest ingestion rate.
Changing \textit{acks} to either \textit{1} or \textit{all} resulted in a rise to a rate of about 320K\,MPS.
Concerning the \textit{batch size}, the increase resulted in a higher ingestion rate for the scenarios without acknowledgments.
Specifically, a rise of more than 20K\,MPS can be observed.
For the other acknowledgment settings, the raised \textit{batch size} led to an unstable ingestion rate.

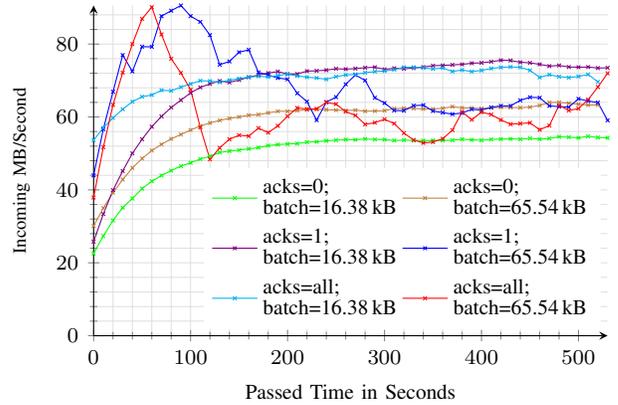
\begin{figure}[!h]
\centering
\begin{tikzpicture}
\begin{axis}[
    xlabel=Passed Time in Seconds,
    ylabel=Incoming MB/Second,
    y label style={at={(axis description cs:0.04,.5)},
    font=\fontsize{7}{7}\selectfont},
    axis y line*=left,
    axis x line*=bottom,
    axis lines = left,
    xmin=0,
    ymin=0,
    grid=both,
    grid style={line width=.1pt, draw=gray!25},
    minor tick num=4,
    width=0.95\columnwidth,
    legend pos=south east,
    height=17em,
    x tick label style={font=\fontsize{8}{8}\selectfont},
    y tick label style={font=\fontsize{8}{8}\selectfont},
    x label style={font=\fontsize{8}{8}\selectfont},
    legend columns=2,
    legend style=
    {at={(0.6,0.02)},
    anchor=south,
    draw=none,
    font=\fontsize{8}{8}\selectfont,
    cells={align=left}
    },
]

\addplot[
color = green,
mark=x,
mark size=1pt
] table [
y expr=\thisrow{Value}/1000000,
x=Time,
col sep=tab
]
{perf_analysis/df-ack0-1m-bytespersec.csv};

\addplot[
color = brown,
mark=x,
mark size=1pt
] table [
y expr=\thisrow{Value}/1000000,
x=Time,
col sep=tab
]
{perf_analysis/df-ack0-b65-1m-bytespersec.csv};

\addplot[
color = violet,
mark=x,
mark size=1pt
] table [
y expr=\thisrow{Value}/1000000,
x=Time,
col sep=tab
]
{perf_analysis/df-ack1-1m-bytespersec.csv};

\addplot[
color = blue,
mark=x,
mark size=1pt
] table [
y expr=\thisrow{Value}/1000000,
x=Time,
col sep=tab
]
{perf_analysis/df-ack1-b65-1m-bytespersec.csv};

\addplot[
color = cyan,
mark=x,
mark size=1pt
] table [
y expr=\thisrow{Value}/1000000,
x=Time,
col sep=tab
]
{perf_analysis/df-ackall-1m-bytespersec.csv};

\addplot[
color = red,
mark=x,
mark size=1pt
] table [
y expr=\thisrow{Value}/1000000,
x=Time,
col sep=tab
]
{perf_analysis/df-ackall-b65-1m-bytespersec.csv};

\legend{
acks=0;\\batch=16.38\,kB,
acks=0;\\batch=65.54\,kB,
acks=1;\\batch=16.38\,kB,
acks=1;\\batch=65.54\,kB,
acks=all;\\batch=16.38\,kB,
acks=all;\\batch=65.54\,kB
}
\end{axis}
\end{tikzpicture}
\caption{Incoming MB/second - one-minute rate, configured 1,000K\,MPS}
\label{fig:distrDataSize1m}
\end{figure}

Figure~\ref{fig:distrDataSize1m} shows the incoming data rate in MB/second, which is provided by Apache Kafka as the metric \textit{BytesInPerSec}.
The chart fits the corresponding ingestion rates shown in  Figure~\ref{fig:1mioack0}.
The highest peaks are at about 90\,MB/second.
The measured maximum network bandwidth between the Apache Kafka brokers is about 117.5\,MB/second, see Table~\ref{tab:system}.
Therefore, if further network traffic is created, that is not captured by the \textit{BytesInPerSec} metric, the bandwidth of the employed commodity network could be a limiting factor in peak situations if data is sent from a remote host.
As we executed the data sender on the node storing the corresponding topic partition, there was intra-node transfer and we used the loopback interface with its higher bandwidth of about 908\,MB/second, which is not a bottleneck.
The determined write performance of about 70\,MB/second described in Table~\ref{tab:system} is even closer to the observed limits in Figure~\ref{fig:distrDataSize1m}.
Depending on how optimized Apache Kafka writes to disk, the achievable performance might be higher.
Nevertheless, the observations lead to the conclusion that the ingestion rate is likely to be disk-bound in the viewed benchmark setting.

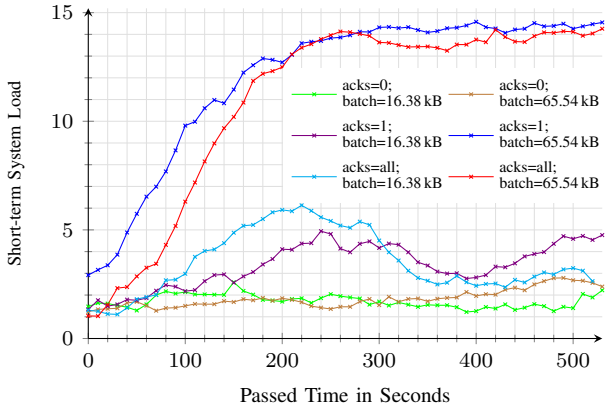
\begin{figure}[!htb]
\centering
\begin{tikzpicture}
\begin{axis}[
    xlabel=Passed Time in Seconds,
    ylabel=Short-term System Load,
    y label style={at={(axis description cs:0.04,.5)},
    font=\fontsize{7}{7}\selectfont},
    axis y line*=left,
    axis x line*=bottom,
    axis lines = left,
    xmin=0,
    ymin=0,
    ymax=15.2,
    grid=both,
    grid style={line width=.1pt, draw=gray!25},
    minor tick num=4,
    width=0.95\columnwidth,
    legend pos=south east,
    height=17em,
    x tick label style={font=\fontsize{8}{8}\selectfont},
    y tick label style={font=\fontsize{8}{8}\selectfont},
    x label style={font=\fontsize{8}{8}\selectfont},
    legend columns=2,
    legend style=
    {at={(0.7,0.41)},
    anchor=south,
    draw=none,
    font=\fontsize{6}{6}\selectfont,
    cells={align=left}
    },
]

\addplot[
color = green,
mark=x,
mark size=1pt
] table [
y expr=\thisrow{Value},
x=Time,
col sep=tab
]
{perf_analysis/df-ack0-1m-loadshortterm.csv};

\addplot[
color = brown,
mark=x,
mark size=1pt
] table [
y expr=\thisrow{Value},
x=Time,
col sep=tab
]
{perf_analysis/df-ack0-b65-1m-loadshortterm.csv};

\addplot[
color = violet,
mark=x,
mark size=1pt
] table [
y expr=\thisrow{Value},
x=Time,
col sep=tab
]
{perf_analysis/df-ack1-1m-loadshortterm.csv};

\addplot[
color = blue,
mark=x,
mark size=1pt
] table [
y expr=\thisrow{Value},
x=Time,
col sep=tab
]
{perf_analysis/df-ack1-65b-1m-loadshortterm.csv};

\addplot[
color = cyan,
mark=x,
mark size=1pt
] table [
y expr=\thisrow{Value},
x=Time,
col sep=tab
]
{perf_analysis/df-ackall-1m-loadshortterm.csv};

\addplot[
color = red,
mark=x,
mark size=1pt
] table [
y expr=\thisrow{Value},
x=Time,
col sep=tab
]
{perf_analysis/df-ackall-65b-1m-loadshortterm.csv};

\legend{
acks=0;\\batch=16.38\,kB,
acks=0;\\batch=65.54\,kB,
acks=1;\\batch=16.38\,kB,
acks=1;\\batch=65.54\,kB,
acks=all;\\batch=16.38\,kB,
acks=all;\\batch=65.54\,kB
}
\end{axis}
\end{tikzpicture}
\caption{Short-term system load of the Apache Kafka broker containing the topic partition - one-minute rate, configured 1,000K\,MPS}
\label{fig:sysload}
\end{figure}

Figure~\ref{fig:sysload} shows the short-term system load of the broker containing the topic partition, which is the server where the data sender is executed.
The system load gives an overview over the CPU and I/O utilization of a server, i.e., also reflecting performance limits regarding disk writes.
It is defined as the number of processes demanding CPU time, specifically processes that are ready to run or waiting for disk I/O.
Figure~\ref{fig:sysload} shows one-minute-averages of this KPI.
As we are using servers with an eight-core CPU each, it is desirable that no node exceeds a system load of eight to do not over-utilize a machine.

Figure~\ref{fig:sysload} reveals that in all settings which led to a steady input rate, the broker node has a system load lower than eight and thus, seems to be not over-utilized from a system load perspective.
The two remaining scenarios show the highest system loads with a value close to 15, which indicates an over-utilization that could limit the achievable ingestion rate.
Interestingly, the system load is not proportional to the corresponding ingestion rates.
At the time of the peak ingestion rate, e.g., the highest system load has not reached its maximum.
That might be an indicator of a growing number of waiting write operations.

\subsection{Input Rate of Overall 500K\,MPS with Two Data Senders}
\begin{figure}[!htb]
\centering
\begin{tikzpicture}
\begin{axis}[
    xlabel=Passed Time in Seconds,
    ylabel=Incoming Messages/Second,
    y label style={at={(axis description cs:0.04,.5)}},
    axis y line*=left,
    axis x line*=bottom,
    axis lines = left,
    xmin=0,
    xmax=620,
    ymin=0,
    ymax=500000,
    grid=both,
    grid style={line width=.1pt, draw=gray!25},
    minor tick num=4,
    width=0.95\columnwidth,
    legend pos=south east,
    x tick label style={font=\fontsize{8}{8}\selectfont},
    x label style={font=\fontsize{8}{8}\selectfont},
    y tick label style={font=\fontsize{8}{8}\selectfont},
    y label style={font=\fontsize{8}{8}\selectfont},
    height=15em,
         legend style=
    {at={(0.65,0.02)},
    anchor=south,
    draw=none,
    font=\fontsize{8}{8}\selectfont
    },
]

\addplot[mark=none, black,dashed, thick, samples=2] coordinates {(0,500000) (615,500000)};

\addplot[
color = blue,
mark=x,
mark size=1pt
] table [
y=Value,
x=Time,
col sep=tab
]
{perf_analysis/df25.csv};

\addplot[
color = green,
mark=x,
mark size=1pt
] table [
y=Value,
x=Time,
col sep=tab
]
{perf_analysis/df26.csv};

\addplot[
color = purple,
mark=x,
mark size=1pt
] table [
y=Value,
x=Time,
col sep=tab
]
{perf_analysis/df27.csv};

\addplot[
color = brown,
mark=x,
mark size=1pt
] table [
y=Value,
x=Time,
col sep=tab
]
{perf_analysis/df28.csv};

\legend{
500K\,MPS,
acks=0; both local,
acks=1; both local,
acks=1; 1 local - 1 remote,
acks=1; both remote (different hosts)
}
\end{axis}
\end{tikzpicture}
\caption{Ingested messages/second - one-minute rate, configured 500K\,MPS in total with two data senders}
\label{fig:distr}
\end{figure}
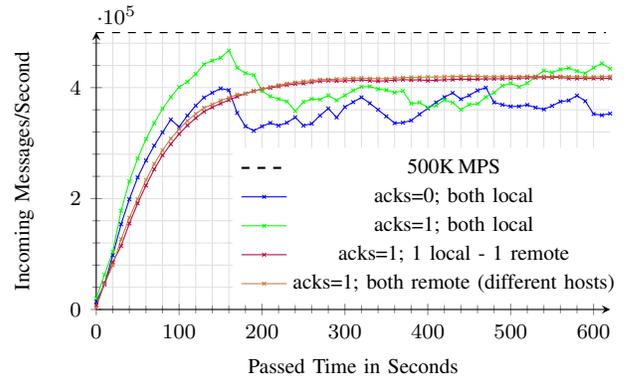

To see if server resources regarding CPU are a limiting factor, we distributed the data sender.
We included the default \textit{batch size} and left out the measurements where \textit{acks} are set to \textit{all} as they are, similar to the previously presented results, practically identical to the runs with \textit{acks} set to \textit{1}.
Figure~\ref{fig:distr} shows the achieved ingestion rates.
The blue and green lines illustrate the runs where both senders run locally, i.e., on the broker node containing the topic partition.
The blue line visualizes the results for disabled \textit{acks} and the green line those for an \textit{acks} level of \textit{1}.
The purple line in Figure~\ref{fig:distr} represents the run where one data sender is invoked on the broker that stores the topic and one data sender at another broker.
The brown line shows the results for the run where the data senders are executed on the two brokers that do not store the topic.

Our measurements show that the two settings where at least one data sender is executed remotely lead to the same result: a steady input rate of about 420K\,MPS.
The benchmark runs having both senders run locally have a different outcome.
Similar to the previous results, the \textit{acks} set to \textit{1} overall outperforms \textit{acks} set to \textit{0}.
However, neither configuration reaches a steady input rate.
Both have the highest spike at the beginning, which is a behavior observed before.

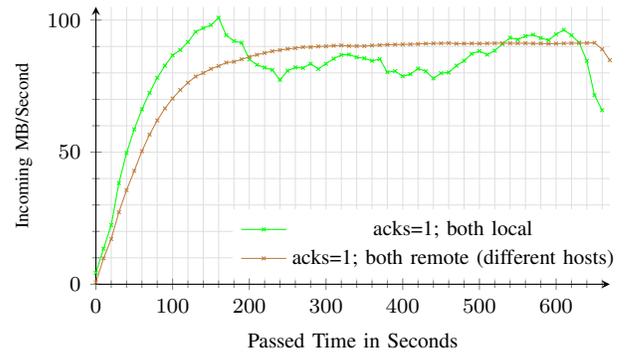
\begin{figure}[]
\centering
\begin{tikzpicture}
\begin{axis}[
    xlabel=Passed Time in Seconds,
    ylabel=Incoming MB/Second,
    y label style={at={(axis description cs:0.04,.5)},
    font=\fontsize{7}{7}\selectfont},
    axis y line*=left,
    axis x line*=bottom,
    axis lines = left,
    xmin=0,
    ymin=0,
    ymax=105,
    grid=both,
    grid style={line width=.1pt, draw=gray!25},
    minor tick num=4,
    width=0.95\columnwidth,
    legend pos=south east,
    height=15em,
    x tick label style={font=\fontsize{8}{8}\selectfont},
    y tick label style={font=\fontsize{8}{8}\selectfont},
    x label style={font=\fontsize{8}{8}\selectfont},
    legend style=
    {at={(0.65,0.02)},
    anchor=south,
    draw=none,
    font=\fontsize{8}{8}\selectfont
    },
]

\addplot[
color = green,
mark=x,
mark size=1pt
] table [
y expr=\thisrow{Value}/1000000,
x=Time,
col sep=tab
]
{perf_analysis/df29.csv};

\addplot[
color = brown,
mark=x,
mark size=1pt
] table [
y expr=\thisrow{Value}/1000000,
x=Time,
col sep=tab
]
{perf_analysis/df30.csv};

\legend{
acks=1; both local,
acks=1; both remote (different hosts)
}
\end{axis}
\end{tikzpicture}
\caption{Incoming MB/second - one-minute rate, configured 500K\,MPS in total with two data senders}
\label{fig:distrDataSize}
\vspace{-1em}
\end{figure}

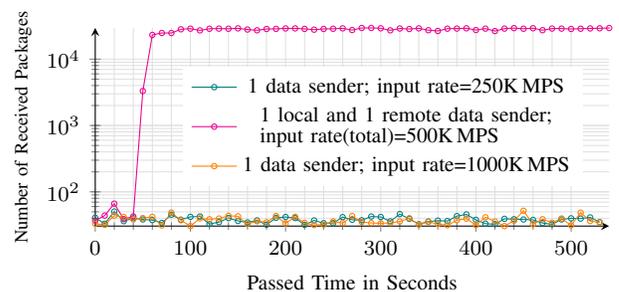
\begin{figure}[]
\centering
\vspace{-1em}
\begin{tikzpicture}
\begin{axis}[
    xlabel=Passed Time in Seconds,
    ylabel=Number of Received Packages,
    y label style={at={(axis description cs:0.04,.5)},
    font=\fontsize{7}{7}\selectfont},
    axis y line*=left,
    axis x line*=bottom,
    axis lines = left,
    xmin=0,
    xmax=540,
    ymin=0,
    ymode=log,
    grid=both,
    grid style={line width=.1pt, draw=gray!25},
    minor tick num=4,
    width=0.95\columnwidth,
    legend pos=south east,
    height=12em,
    x tick label style={font=\fontsize{8}{8}\selectfont},
    y tick label style={font=\fontsize{8}{8}\selectfont},
    x label style={font=\fontsize{8}{8}\selectfont},
    legend style=
    {at={(0.56,0.2)},
    anchor=south,
    draw=none,
    font=\fontsize{8}{8}\selectfont,
    cells={align=left}
    },
]

\addplot[
color = teal,
mark=o,
mark size=1pt
] table [
y expr=\thisrow{Value},
x=Time,
col sep=tab
]
{perf_analysis/df-ack1-250k-eth0.csv};

\addplot[
color = magenta,
mark=o,
mark size=1pt
] table [
y expr=\thisrow{Value},
x=Time,
col sep=tab
]
{perf_analysis/df-ack1-dist-eth0.csv};

\addplot[
color = orange,
mark=o,
mark size=1pt
] table [
y expr=\thisrow{Value},
x=Time,
col sep=tab
]
{perf_analysis/df-ack1-1M-eth0.csv};

\legend{
1 data sender; input rate=250K\,MPS,
1 local and 1 remote data sender;\\input rate(total)=500K\,MPS,
1 data sender; input rate=1000K\,MPS
}
\end{axis}
\end{tikzpicture}
\caption{Packages received/second on interface eth0 - one-minute rate, \textit{acks=1}}
\label{fig:eth0}
\end{figure}

Next to almost identical trends regarding the ingestion rates compared to the two benchmark settings presented before, the system loads are also equal with a value consistently approaching 15.
This again indicates an over-utilization of the server.
The system load never exceeds a value of three on any server with distributed data senders.
Nevertheless, the input rates for the setting with two data senders are the overall highest on average, with a maximum input rate of about 460K\,MPS.
Figure~\ref{fig:distrDataSize} shows the data size characteristics.
The amount of incoming data in MB/second is visualized for the setting where both data senders were executed locally and remotely with \textit{acks} set to \textit{1}.
The maximally achieved input rate of Figure~\ref{fig:distr} corresponds to an input rate of about 100\,MB/s.
For the constant input rate where both senders were executed remotely, a size-wise input of close to 92\,MB/s is reached.
The amount of incoming MB/s exceeds the measured maximum write performance mentioned in Table~\ref{tab:system}, which could be due to increased parallelization or an optimized way of storing messages implemented in Apache Kafka.
As the fully distributed setting uses the \textit{eth0} interface to the broker, the network bandwidth of about 117.5\,MB/second applies.
Since the reported number of incoming bytes is close to this limit and metadata or further traffic might not be captured, the network represents a potential bottleneck.

Figure~\ref{fig:eth0} visualizes the number of packages received on interface \textit{eth0} exemplary for three benchmark runs with a logarithmic scale on the y-axis.
This interface is the only one next to the loopback interface on the used servers.
The figure highlights the differences caused by changes in data sender locality.
While the number of received packages is not impacted if data is only sent from the node where the corresponding target topic is stored, transmitting data from a remote host significantly increases this KPI.
Specifically, no remote data senders result in between 25 and 60 received packages on \textit{eth0}.
One remote data sender amounts to about 30K received packages, while two remote data senders are about to double this number accordingly.

\subsection{Summary}
Our benchmark results reveal two main insights:
Firstly, although a single data sender can create an input rate of 250K\,MPS as shown in Figure~\ref{fig:250k}, two independently executed data senders do not reach the expected input rate of 500K\,MPS.
Secondly, we show the influence of where data senders are invoked.
When two data senders are executed in parallel on the same host, they are able to overwhelm the server or impede each other, as the observed system load of about 15 indicates.
Another limiting factor can be found in the write-to-disk performance of the used server and the network bandwidth when sending data from a remote host.
The observed memory usage was never close to its limits for any of the presented benchmark scenarios.

The most promising configuration in the study, which led to stable input rates, has the default \textit{batch size} and \textit{acks} set to \textit{1} or \textit{all}.
A stable rate is desirable as it leads to predictable system behavior.
Multiple data senders distributed across nodes are able to increase the achievable ingestion rates.
An input rate of about 250K\,MPS to Apache Kafka can be achieved using a single data sender.

\section{Related Work}
\label{chap:rw}
Dobbelaere and Esmaili~\cite{DBLP:conf/debs/DobbelaereE17} compare Apache Kafka with RabbitMQ~\cite{rabbitmq}, another open-source message broker.
In their work, they compare both solutions qualitatively and quantitatively.
The impact of different acknowledgment levels is one of the factors the authors evaluated in their study.
However, their results do not show a clear difference in the achieved throughput between an \textit{acks} level of one and zero in the analyzed setting.

Noac’h, Costan, and Bougé~\cite{DBLP:conf/bigdataconf/NoachCB17} evaluate the performance of Apache Kafka in combination with stream processing systems.
They also study the influence of Apache Kafka characteristics, the producer batch size being one of them.
Similar to our results, their findings reveal that an increased batch size does not necessarily lead to a higher throughput.

Kreps et al.~\cite{kreps2011kafka} present a performance analysis of three systems:  Apache Kafka, RabbitMQ, and Apache ActiveMQ~\cite{activemq}.
Similar to the work presented before, they analyze the influence of the batch size of the Apache Kafka producer.
Next to the producer, they study the Apache Kafka consumer behavior and compare it to the other systems.
The achieved throughput for Apache Kafka in~\cite{kreps2011kafka} is in a similar range as the results of this paper.

Apache Pulsar~\cite{pulsar} is a message broker originally developed at Yahoo!.
It makes use of the distributed storage service Apache BookKeeper~\cite{bookkeeper}.
Similarly to Apache Kafka, Apache Pulsar employs the concept of topics to which producers can send messages and to which consumers can subscribe.
The blog post~\cite{pulsar} presents a brief performance analysis.
The throughput that was achieved in their study using an SSD is 1,800K\,MPS.
However, they do not give details about the test setup, making it hard to assess the results.

Next to these open-source systems, there are commercial solutions, such as Google Cloud Pub/Sub~\cite{googlepubsub} and Amazon Simple Queue Service (SQS)~\cite{amazonsqs}.
Studying further systems were out of the scope of this paper.

\section{Lessons Learned}
\label{sec:lessonslearned}

Overall, the conducted performance study shows that estimates regarding the performance impact of different Apache Kafka producer configurations, based on experiences and perceptions, are not always true.
We particularly emphasize two unexpected behaviors that are present in the collected results.
These findings are related to two configuration options of the Kafka producer: the acknowledgments level and the batch size.
We theorized that a lower level of acknowledgments would necessarily lead to a higher input rate, as sending of messages and waiting for an acknowledgment of their arrival represents an overhead.
However, our study shows that this theory does not hold in all observed cases.

Similarly, this is true for the Kafka producer batch size configuration option.
Our expectation was that increasing the batch size would lead to a higher input rate, since the number of send actions can be reduced, also lowering the overall overhead.
The presented performance study revealed that higher batch sizes can indeed increase the input rate as expected, see Figure~\ref{fig:1mioack0}.
However, although increasing the batch size may lead to a higher peak input rate, it often caused a non-steady, fluctuating input rate.
Additionally, the observed average input rates for \textit{acks} set to \textit{1} and \textit{all} are lower for the configuration with an increased batch size.

As a result of these experiences, we highlight the importance of benchmarking message brokers to explore their behavior in application scenarios and to obtain realistic KPI's.
This allows basing discussions regarding technology selection on facts.
Additionally, it lowers the likelihood of wrong assessments.

Regarding the tooling utilized for observing the performance characteristics of Apache Kafka, we made use of virtualization technology in the form of Docker containers.
This turned out to be a low-effort way of deploying different systems.
Also, orchestrating these independently running containers, as in the proposed benchmarking architecture, did not introduce any significant management overhead.
So for similar settings, we recommend using virtualization technologies due to the easy and fast setup.
Additionally, moving to a different server or updating systems are tasks that can be done with low effort.

\section{Conclusion and Future Work}
\label{sec:conclusion}

We propose and implement a monitoring architecture for Apache Kafka and similar systems running in a JVM.
We incorporate state-of-the-art technologies such as \textit{Grafana} and \textit{collectd} with a focus on ease of use and adaptability for future measurements.
We performed a performance study of Apache Kafka using our developed benchmarking setup.
We evaluate and discuss our benchmark results for varying data sender configurations.
The benchmark artifacts, such as the data sender tool and the \textit{Grafana} dashboard, are published for transparency and reproducibility.

In the configuration featuring a single topic with one partition and a replication factor of one, we achieve a maximum steady ingestion rate to Apache Kafka of about 420K\,MPS or about 92\,MB/s.
We quantified the impact of the Apache Kafka producer batch size, acknowledgment level, data sender locality as well as of additional aspects on the input rate performance.
We analyzed the server's behavior during the benchmark runs to explore potential performance bottlenecks.
Our study highlights the influence of the chosen acknowledgment level.
Configurations with enabled acknowledgments showed better performance, i.e., a higher message input rate, which was counter to our working hypothesis.
An analysis of why this behavior was observed is part of future research.
Moreover, a comparison of apache Kafka to similar systems, such as Apache Pulsar or RabbitMQ, would be of interest to the research community.
Further work should focus on the analysis of Apache Kafka producers employed in data stream processing frameworks, such as Apache Flink or Apache Beam.
These systems often provide their own Apache Kafka producer implementations or interfaces.
It would be interesting to investigate if these embedded producers perform differently in comparable settings regarding the achievable input rate.

Furthermore, it is valuable to know how the input rate behaves when scaling via, e.g., the number of topic partitions.
With a growing number of partitions, scaling the number of broker nodes becomes an additional dimension whose influence can be measured.
The impact of higher replication factors is another open domain of future research.

\bibliographystyle{IEEEtran}
\bibliography{sample-base.bib}

\end{document}